\documentclass{article}
\usepackage{amssymb}

\def\bd{\begin{definition}}
\newtheorem{lemma}{Lemma}[section]
\newtheorem{remark}{Remark}[section]
\newtheorem{proposition}{Proposition}[section]
\newtheorem{definition}{Definition}[section]
\newtheorem{theorem}{Theorem}[section]

\def\ed{\end{definition}}
\def\bt{\begin{theorem}}
\def\et{\end{theorem}}
\def\be{\begin{equation}}
\def\ee{\end{equation}}

\def\ee{\end{equation}}
\def\bes{$$}
\def\ees{$$}
\def\ri{\right}
\def\pa{\partial}
\def\bea{\begin{eqnarray}}
\def\eea{\end{eqnarray}}
\def\beas{\begin{eqnarray*}}
\def\eeas{\end{eqnarray*}}

\begin{document}

\baselineskip 16pt plus 1pt minus 1pt
\begin{titlepage}
\begin{flushright}
\end{flushright}
\vspace{0.2cm}
\begin{center}
\begin{Large}
%\fontfamily{cmss}
%\fontsize{17pt}{27pt}
%\selectfont
\textbf{  Massless scalar field in two--dimensional de Sitter universe}
\end{Large}\\
{Dedicated to Jacques Bros for his birthday}\\
\bigskip
\begin{large} {Marco
Bertola}$^{\dagger}$\footnote{Work supported in part by the Natural
    Sciences and Engineering Research Council of Canada (NSERC),
    Grant. No. 261229-03 and by the Fonds FCAR du
    Qu\'ebec No. 88353.}, Francesco Corbetta$^\ddagger$, Ugo Moshella$^{\ddagger\sharp}$
\end{large}
\\
\bigskip
\begin{small}
$^{\dagger}$ {\em Centre de recherches math\'ematiques,
Universit\'e de Montr\'eal\\ C.~P.~6128, succ. centre ville, Montr\'eal,
Qu\'ebec, Canada H3C 3J}  and \\
{\em Department of Mathematics and
Statistics, Concordia University\\ 7141 Sherbrooke W., Montr\'eal, Qu\'ebec,
Canada H4B 1R6} \\
\smallskip 
$^\ddagger$ {\em Dipartimento di fisica e Matematica, Universit\`a dell'Insubria,
22100 Como, (Italia)} \\
\smallskip 
$^\sharp$ {\em INFN, Sez. di Milano, Via Celoria 2,
Milano (Italia)}
\end{small}
\end{center}
\bigskip
%%%%%%%%%%%%%%%%  Abstract  %%%%%%%%%%%%%%%%
\begin{center}{\bf Abstract}\\
\end{center}
We study the massless minimally coupled scalar field on a two--dimensional de
Sitter space--time in the setting of
axiomatic quantum field theory. We construct the
invariant Wightman distribution obtained as the
renormalized zero--mass limit of the massive
one. Insisting on gauge invariance of the model we construct a vacuum
state and a Hilbert space of physical states which
are invariant under the action of the whole de Sitter group.
We also present the integral expression of the conserved charge which
generates the gauge invariance and propose a definition of dual field.
\medskip
%\begin{small}
%\end{small}	
\bigskip
\bigskip
\bigskip
\bigskip

\end{titlepage}

\section{de Sitter geometry and  the massive scalar
field}

The de Sitter space--time may be represented as a $d$--dimensional
one--sheeted hyperboloid 
\be 
X_d(R)=X_d=\left\{x\in{\mathbb
R}^{d+1}\ :\ x^{(0)^2}-x^{(1)^2}-\cdots-x^{(d)^2}=-R^2\ri\} 
\ee
embedded in a Minkowski ambient space ${\mathbb R}^{d+1}$ with
scalar product $x\cdot
y=x^{(0)}y^{(0)}-x^{(1)}y^{(1)}-\cdots-x^{(d)}y^{(d)}$ (and
$x^2=x\cdot x$).
 The manifold
$X_d$ is then equipped with a causal ordering relation induced by that of
${\mathbb R}^{d+1}$.
 The invariance  group of the de Sitter space--time $G_d=SO_0(1,d)$.
The $G_d$--invariant volume form on $X_d$ will be denoted by ${\rm
d}\sigma(x)$. Let us introduce the complex hyperboloid 
\be
X^{(c)}_d(R)=X^{(c)}_d=\left\{z=x+iy\in{\mathbb C}^{d+1}\ :\
z^{(0)^2}-z^{(1)^2}-\cdots -z^{(d)^2}=-R^2\ri\}\ , 
\ee 
equivalently
characterized as the set
$$ 
X_d^{(c)}=\left\{(x,y)\in{\mathbb
R}^{d+1}\times  {\mathbb R}^{d+1}\ :\ x^2-y^2=-R^2,\ x\cdot
y=0\ri\}\ .
$$
 We define the  following open subset of $X_d^{(c)}$:
 \be
{\mathcal T}^+=T^+\cap X_d^{(c)},\qquad {\mathcal T}^-=T^-\cap
X_d^{(c)}\ , 
\ee
 where $T^\pm={\mathbb R}^{d+1}+i V^\pm$ are the
so-called {\em forward} and {\em backward tubes} in ${\mathbb
C}^{d+1}$ \cite{[BM]} (here $V_\pm$ are the standard future/past
lightcones in the ambient Minkowskian manifold). These  are the
(minimal) analyticity domains of the Fourier--Laplace transforms of
tempered distributions $\widetilde{f}(p)$ with support contained in
$\overline{V^+}$ or in $\overline{V^-}$ .

The r\^ole of Wightman axioms (\cite{[SW]}) for a QFT in Minkowski are replaced by
other axioms of similar nature  (\cite{[BGM]},\cite{[BM]} and
\cite{[M]}) where the spectral condition is replaced by the notion of
minimal analyticity. Specifically the axioms to be satisfied are (for
a quasi-free theory):
\begin{enumerate}
\item Positivity: $\displaystyle \int_{X_d\times X_d}\hspace{-30pt}\mathcal W(x,x')
  f(x) \overline f(x') {\rm d}\sigma(x) {\rm d}\sigma(x') \geq 0,\
  \forall \ f\in \mathcal D(X_d)$.
\item Locality: ${\mathcal W}(x,x')={\mathcal W}(x',x)$ for every
space--like separated pair
$(x,x')$;
\item Covariance: ${\mathcal W}(gx,gx')={\mathcal W}(x,x'),\ \forall g\in G_d$
\item Normal analyticity: the two point function ${\mathcal W}(x,x')=
\langle\Omega,\phi(x),\phi(x')\Omega\rangle$ is the boundary value of
a function which is
analytic in the domain ${\mathcal T}_{12}$ of $X_d^{(C)}\times X_d^{(c)}$.
\end{enumerate}
For a
 minimally coupled
Klein--Gordon the  field equation may be written as 
\be
 \square
\phi+\mu^2\phi=0, 
\ee 
where  $\mu^2=M^2+\zeta\rho$ is the relevant
mass parameter, and  $\zeta$ is the coupling constant to the scalar
curvature $\rho$. A plane-wave analysis similar in spirit to the
Minkowskian case and subsequent evaluation of integrals \cite{[BM]}
lead to the following two-point function for such a scalar field 
\be
W_\nu(z,z')=\frac{2c_{d,\nu}e^{\pi\nu}\pi^{d/2}}{R^{d-1}\Gamma(d/2)}P^{(d+1)}_{-\frac{d-1}{2}
+i\nu}(\lambda)\ ,\,\,\, \lambda := \frac {z\cdot z'}{R^2} \,\,\,\,
 R^2 \mu^2 = \frac {(d-1)^2}4 + \nu^2
\label{equ:twopointm} 
\ee
 with 
 \bea &&
P^{(d+1)}_{-\frac{d-1}{2}+i\nu}(\lambda)=\frac{\Gamma(d/2)}{\sqrt\pi
\Gamma\left(\frac{d-1}{2}\ri)}\int_0^\pi\left(\lambda+\sqrt{\lambda^2-1}\cos\theta\ri)^{-\frac{d-1}{2}+i\nu}\sin^{d-2}\theta\
{\rm d}\theta\ ,\nonumber \\
&&
c_{d,\nu}=\frac{\Gamma\left(\frac{d-1}{2}+i\nu\right)\Gamma\left(\frac{d-1}{2}-i\nu\right)
e^{-\pi\nu}}{2^{d+1}\pi^d}\ . 
\eea This two-point function gives
rise to a QFT satisfying all the previously mentioned axioms and can
be shown to be unique.
\section{Massless limit}
\label{sec:zerolimit}
We
 are interested in the two--dimensional ($d=1$)  massless
limit $\mu^2\to 0$ for the two--point function in Eq. \ref{equ:twopointm}
namely the Wightman function of a massless minimally coupled
scalar field.\\
As the Minkowskian case shows \cite{[S]}  \cite{[MPS]} \cite{[MPS2]}, two--dimensional massless QFTs present us
with infrared singularities. In the present context -being de Sitter
 compact in the spatial directions- we may
expect a natural IR cutoff. This indeed  will be the case  at the
price of  violating  the equation of motion
{\em and}  the positivity assumption.
We will show that restoring the positivity by means of Krein topology
automatically restores
 the {\em effective} equation of motion. This means that the equation
of motion will be
 automatically  satisfied
on the  physical states ({\em \'a\ la} Gupta-Bleuer).\par
The Krein construction
ensures that the Hilbert space obtained by closing the space of
local states is a ``maximal''  space
%in the sense that
% every Hilbert space obtained as closure of ${\mathcal
%D}(X_2)$ with a topology majorizing the indefinite inner product is contained in this space
(\cite{[S]}).

Let us proceed and perform the massless limit on the Wightman two--point
function. We can write the
massive two--point function as
 \be
W_\nu(\lambda)=\frac{\Gamma\left(1-\alpha\ri)\Gamma(\alpha)}{4\pi}F\left(1-\alpha,\alpha;1;
\frac{1-\lambda}{2}\ri)\ , 
\ee
 where $\alpha=\frac{1}{2}-i\nu$, $F(a,b;c;x)$ is the
hypergeometric function and $\lambda=\frac{z\cdot z'}{R^2}$.
 For $\alpha\to 0$ (or $\alpha\to 1$),
we have that $\mu^2 = 1/R^2\, \alpha(1-\alpha)\to  0$.\\
The gamma function $\Gamma(\alpha)$ has a simple pole at the origin
so that we have to renormalize the massive two--point function in
order to obtain a finite result. To this end
we expand $W_\nu(\lambda)$ in power series of $\alpha$:
 \bea
W_\nu(\lambda)&=&\frac{\Gamma\left(1-\alpha\ri)\Gamma(\alpha)}{4\pi}+
\frac{\Gamma\left(1-\alpha\ri)\Gamma(\alpha)\alpha}{4\pi}\sum_{n=1}^\infty
\frac{1}{n}\left(\frac{1-\lambda}{2}\ri)^n+o(1) 
\eea
  By subtraction of the divergence due to the first term  we obtain
\be 
W_0\equiv W_0(\lambda)\equiv
\lim_{\alpha\to0}\left[W_\alpha(\lambda)-
\frac{\Gamma\left(1-\alpha\ri)\Gamma(\alpha)}{4\pi}
\ri]=-\frac{1}{4\pi}\ln\left[ -\frac{\left(z-z'\ri)^2}{R^2}\ri]. 
\ee
Note that this is the same function as in the flat case, but with
the argument being the (hyperbolic for time--like separation) sine
of the pseudo--distance  in the de Sitter manifold.
%In
%order to simplify formul\ae, in the following we will put $R=1$.

Such a  two--point function is analytic in the extended tube
 and invariant under the action of
the complex de Sitter group $G^{(c)}_2$ \cite{[BM]}.
 The boundary value $\mathcal{W}_0$ of $W_0$ is therefore
invariant
for a de Sitter transformation $g$, it is local and analytic in the domain
$\mathcal{T}_{12}$.\par
This renormalization procedure has however introduced an anomaly in the equation
 of motion; indeed taking the limit
\bea
 && \square \mathcal W_0(x,x') =
 \lim_{\alpha\to
 0} \square\left(\mathcal{W}_\alpha(x,x')-C_{2,\alpha}\ri)
 = - \lim_{\alpha \to 0} \mu^2\mathcal{W}_\alpha(x,x') = \nonumber \\ && =- \lim_{\alpha
 \to 0} \mu^2  C_{2,\alpha} = -\frac 1 {4\pi{\rm R}^2}
\eea 
we get the equation of motion 
\be
\square\mathcal{W}_0(x,x')=-\frac{1}{4\pi{\mathrm R}^2}. 
\ee
 The
renormalization procedure has introduced an
 {\em anomaly} in the equation of motion. Moreover the resulting
 kernel is {\bf not positive} essentially due to the logarithmic
 singularity at short distances, another drawback of this
 renormalization procedure.\\
This forces us to use the extended
 Wightman axioms which require the introduction of a Krein topology,
  restoring at the same time positivity and equation of motion
 on the physical states.
%For this reason this two--point function is not the v.e.v. of a
% Klein-Gordon massless scalar field. In order to
%obtain the Wightman distribution of such a field we must add to $\mathcal{W}_0(x,x')$ a function
%$\overline{\chi}(x)+\chi(x')$ such that
%\be
%\square \chi=\frac{1}{4\pi}.
%\ee
%This necessarily breaks de Sitter invariance because the only scalar function over the de
%Sitter hyperboloid which satisfies
%\be
%\chi(x)=\chi(gx),\qquad \forall g\in G_2
%\ee
%is a constant (\cite{[Wein]}).\\
\begin{remark}
One may ask why the solution of the equation $\square F(\lambda)=0$,
namely 
\be
 F(\lambda)=\ln\left(\frac{1-\lambda}{1+\lambda}\ri)\ ,
\ee
 is not an acceptable two point function ($\lambda$ being the
invariant $\frac{z\cdot z'}{R^2}$). The reason is that this function
is invariant under the action of $G_2$ but it is not analytic in the
full extended domain: it is analytic in $\mathbb{C}^2$ minus the
cuts $(-\infty,-1)$ and $(1,\infty)$. In particular the cut
$(1,\infty)$ violates locality for antipodal points on the de Sitter
manifold.
\end{remark}
\section{Invariant Krein space}
\label{sec:krein}
A two--point function which is invariant, analytic and causal, but not positive, allows to
construct, via a GNS construction, a linear space invariant under de Sitter transformation
with an indefinite inner product.
Were the two--point (semi-)positive it would define a
pre-Hilbert structure, and the space of physical states would be
identified with the closure under such norm.
 If the Wightman functions fail to be positive, such identification is
not natural and, in general, not unique.
We have to look for a {\em minimal} Hilbert topology,
{\em i.e.} such that the closure of the span of local states  is
maximal ({\em Krein topology}).
We give here a direct construction tailored to our specific
case.

The invariant two--point function $\mathcal{W}_0$ is positive on the
subspace ${\mathcal D}_0$, the space of null integral test functions
\be 
\mathcal D_0:= \left\{f\in\mathcal D(X_2):\ \int_{X_2}\!\!{\rm
d}\sigma(x)f(x) = 0\ri\} , 
\ee
 because on this subspace the
renormalization procedure by which we obtained the two-point
function is irrelevant. Let us choose and fix a real function
$h\in{\mathcal D}(X_2)$ such that \bea \int_{X_2}{\rm
d}\sigma(x)h(x)=1\ ,\label{equ:h1}\ \qquad \langle h,h\rangle=0\ .
\eea Then, for any   $f\in \mathcal D(X_2)$ we can decompose as
\[f(x)=f_0(x)+\left(\int_{X_2}{\rm d}\sigma(y)f(y)\ri)h(x),\] and
therefore the test function space ${\mathcal D}(X_2)$ is decomposed
in the direct sum 
\be 
{\mathcal D}(X_2)={\mathcal D}_0+H, 
\ee
where $H$ is the one-dimensional space generated by $h$.\\
Following this decomposition the indefinite sesquilinear form given
by $\mathcal W_0$ is written as 
\be
 \langle f,g\rangle=\langle
f_0,g_0\rangle+\left(\int_{X_2}{\rm d}\sigma(x)\overline{f}(x)\ri)
\langle h,g\rangle+\left(\int_{X_2}{\rm d}\sigma(x)g(x)\ri)\langle
f,h\rangle. 
\ee
 The Krein inner product on ${\mathcal D}(X_2)$ is
then defined by 
\be
 (f,g)=\langle f_0,g_0\rangle+\langle
f,h\rangle\langle h,g\rangle+\int_{X_2}{\rm d}\sigma(x)
\overline{f}(x)\int_{X_2}{\rm d}\sigma(y)g(y).%\label{equ:kreininv}
\label{equ:kreinprod} 
\ee
 The inner product (\ref {equ:kreinprod})
is positive semidefinite and hence defines a structure of
pre-Hilbert space. The nihilspace of this pairing is the subspace
\be
 {\mathcal I}_h=\left\{f\in{\mathcal D}_0\ :\ \langle
f_0,f_0\rangle=0,\ \langle h,f\rangle=0\ri\}\ .\label{ideal} 
\ee
 By
quotienting $\mathcal D(X_2)$ and then completing in the ensuing
Hilbert topology we obtain a Hilbert space which we denote with
 ${\mathcal K}^{(1)}$.
\begin{lemma}
The linear functional on ${\mathcal K}^{(1)}$ defined by
\be
F_h(f)=\langle h,f\rangle,
\ee
(where $h$ is the previously fixed test function with unit integral)
has norm equal to one and therefore it defines a normalized element $v_0$ of
${\mathcal K}^{(1)}$, such that $(v_0,f)=F_h(f)$.\\
Furthermore, for all $f\in{\mathcal D}
(X_2)$,
\be
\langle v_0,f\rangle=\int_{X_2}{\rm d}\sigma(x)f(x)\ .
\ee
\end{lemma}
{\bf Proof} Let ${\mathcal I}_h$ be the ideal of the test functions of
zero-norm (\ref{ideal}).
Then  ${\mathcal K}^{(1)}=\overline{{\mathcal D}(X_2)/{\mathcal I}_h}^{\|\cdot\|}$.
Defining
\be
v_0\equiv -4\pi{\mathrm R}^2\square h
\ee
we have
\be
(v_0,f)=\langle v_0,f_0\rangle+\langle v_0,h\rangle\langle
h,f\rangle+\int_{X_2}{\rm d}\sigma(x)v_0(x)\int_{X_2}{\rm
  d}\sigma(y)f(y)=\langle h,f\rangle\ ,
\label{ortho}
\ee
since  $\langle v_0,h\rangle=1$ and $v_0\in{\mathcal D}_0$. Thus $v_0$
is the element
identified by the functional $F_h$ as in the statement of the lemma.\\
We now prove uniqueness of the vector $v_0$. Consider
a different real function $h'$ in ${\mathcal D}(X_2)$
 satisfying eqs. (\ref{equ:h1}). Then
the difference  $h-h'$ belongs to ${\mathcal D}_0$ and hence
 $\square(h-h')$ belongs to the ideal  ${\mathcal I}_h$, so that the
corresponding vectors $v_0'$ and $v_0$ are the same in the quotient space.\\
Finally the norm of $v_0$ is given by 
\be
 (v_0,v_0)=16\pi^2 {\rm
R}^4 \left|\langle\square h,h\rangle\ri|^2=1 
\ee
 and thus
$\|F_h\|=1$. Moreover, we have that 
\be
 \langle v_0,v_0\rangle=0 
 \ee
and, for every function $f$ in ${\mathcal D}(X_2)$, 
\be
 \langle
v_0,f\rangle=-4\pi{\rm R}^2 \langle \square
h,f\rangle=\int_{X_2}{\rm d}\sigma(x)f(x). 
\ee
 Q.E.D.\par
\begin{remark}
 The state $v_0$ plays the same role as
the infinitely delocalized state of \cite{[MPS]}. However
there such state is a limit of test functions and belongs only to the
Krein completion of the space: here on the contrary the state is a
perfectly well-behaved test function obtained by the application of
the Laplace--Beltrami operator $\square$ to the chosen $h$. This may seem
not to be an invariant state, however -as it is proven in
Prop. \ref{prop42}- the
action of an isometry of the space-time does not change its {\em
equivalence class} modulo the ideal ${\mathcal I}_h$.
\end{remark}
\begin{proposition}
The Hilbert space ${\mathcal K}^{(1)}$ is a Krein space and can be written as a direct sum:
\be
{\mathcal K}^{(1)}=\overline{{\mathcal D}_0^\bot/\mathcal I_h}^{\langle\cdot,\cdot\rangle}\oplus V_0\oplus H\ ,
\label{equ:kreindec}
\ee
where ${\mathcal D}_0^\bot$ is the subspace of ${\mathcal D}_0$
orthogonal to (the eq. class of) $v_0$ spanning
$V_0=(\{\lambda v_0\ :\ \lambda\in{\mathbb C}\} +\mathcal
I_h)/\mathcal I_h$ and $H=(\{\lambda h\ :\ \lambda\in{\mathbb C}\} +\mathcal
I_h)/\mathcal I_h$.\\
The metric operator $\eta^{(1)}$ defined by $\langle\cdot,\cdot\rangle=(\cdot,\eta^{(1)}\cdot)$ is
given by
\bea
\eta^{(1)}_{|_{\overline{{\mathcal
D}_0^\bot}^{\langle\cdot,\cdot\rangle}}}
={\mathbf 1}_{|_{\overline{{\mathcal
    D}_0^\bot}^{\langle\cdot,\cdot\rangle}}} \ ,\ \
\eta^{(1)}h=v_0\ ,\ \
\eta^{(1)}v_0=h\ .
\eea
{\em Note}: from now on we will use the same symbols $h,v_0$ to denote
the functions and the equivalence classes in the quotient space.
\end{proposition}
{\bf Proof}
We first note that $\mathcal I_h$ is entirely contained in ${\mathcal
  D}_0^\bot$ since  for any $f\in \mathcal D_0$
\be
 \left(f,v_0\ri) \propto  \langle f,h\rangle 
 \ee
  and hence the
orthogonality to $v_0$ is one of the defining properties of the
vector space $\mathcal I_h$ (see eq. \ref{ideal}).

Let us write the subspace of zero-integral test functions as
${\mathcal D}_0={\mathcal D}_0^\bot\oplus\mathbb C  \{v_0\}$, where the orthogonality is with respect to the
(pre-)Hilbert product $(\cdot,\cdot)$. If restricted to the space ${\mathcal D}_0^\bot$, the Krein product is equal
to the indefinite sesquilinear form $\langle\cdot,\cdot\rangle$. Indeed, for every pair of
test functions $f,g\in{\mathcal D}_0^\bot$ we have
\be
(f,g)=\langle f,g\rangle+\langle f,h\rangle\langle h,g\rangle=\langle f,g\rangle+(f,v_0)(v_0,g)=\langle f,g\rangle\ .
\ee
The mono-dimensional space $H$ is orthogonal to $V_0$ because $(v_0,h)=\langle h,h\rangle=0$.
Moreover we have ${\mathcal D}_0^\bot \bot H$:
\be
(h,f)=\int_{X_2}{\rm d}\sigma(x)h(x)\int_{X_2}{\rm d}\sigma(y)f(y)=0\ .
\ee
This is true also for the Hilbert space $\overline{{\mathcal D}_0^\bot}^{\langle\cdot,\cdot\rangle}$
and thus we have proved the decomposition (\ref{equ:kreindec}). The definition of the metric
operator $\eta^{(1)}$ is now obvious. The Hilbert space ${\mathcal K}^{(1)}$ is then a Krein
space and, being $\eta^{(1)^{-1}}$ continuous, the Hilbert majorant topology defined by the
Krein norm is minimal and the Krein space is maximal.\\
Q.E.D.

The de Sitter group acts naturally on $\mathcal D(X_2)$ by
 $\alpha_g f(x)=f(g^{-1}x)$ and the restriction to
$\overline{{\mathcal D}_0^\bot}^{\langle\cdot,\cdot\rangle}$ is
 unitary since ${\mathcal W}_0$
is invariant. Moreover we have
\begin{proposition}
\label{prop42}
The representation $U(g)$ of the de Sitter group has a unique continuous extension from
${\mathcal D}(X_2)$ to ${\mathcal K}^{(1)}$ and
\be
U(g)v_0=v_0\ .
\ee
\end{proposition}
{\bf Proof.} The only non-obvious part is the invariance of $v_0$. For any
 element $g$ of $SO_0(1,2)$ the function $ \square h(x)-\square h(gx) $
belongs to
${\mathcal I}_h$ (note that $h(x)-h(gx)$ is a function in ${\mathcal D}_0$) and
thus it is zero in ${\mathcal K}^{(1)}$, thus proving that  $U(g)v_0=v_0$. Q.E.D.\par \vskip 4pt
>From the decomposition (\ref{equ:kreindec}) follows that $v_0$ is the unique invariant vector in
${\mathcal K}^{(1)}$.\\
>From the construction of the one--particle Hilbert-Krein space
$\mathcal K^{(1)}$ one can build as usual a Fock--Krein space $\mathcal
K$ and extend there the metric operator $\eta$. Then we can prove
\begin{proposition}
The space ${\mathcal K}$ contains an infinite--dimensional subspace ${\mathcal V}_0$ of
vectors invariant under the de Sitter transformation. However, the
vacuum is still {\em essentially}
unique ({\it i.e.} any strictly positive subspace of invariant vectors is one dimensional).
\end{proposition}
{\bf Proof}. The proof of this theorem follows \cite{[MPS]}
Q.E.D.
\section{Equation of motion and gauge invariance}
In this section we investigate the equation of motion and the
existence of a charge operator generating gauge transformations.\\
As we have remarked above the most important difference between the
massless scalar field  in Minkowski space--time and the covariant
massless scalar field in de Sitter space--time  is that the
invariant state $v_0$ now belongs to the space of test functions
${\mathcal D}(X)$. Therefore  the operator $\phi(v_0)$ belongs to
the field algebra and its introduction does not require any
extension of the latter. This  invariant operator commutes with any
other operator of ${\mathcal A}$: 
\be 
\left[\phi(v_0),\phi(f)\ri]=0,
\ee
 as follows from the explicit expression of $v_0=-4\pi{\rm R}^2
\square h$ and the equations of motions $\square_{x} { W}(x,x') =
\square_{x'}
 { W}(x,x')=- \frac 1 {4\pi{\rm R}^2}$.\par
We begin  by  proving  that the equation of motion for the
two--point function, $\square \mathcal W = -1/(4\pi\mathrm R^2)$
translates into the following equation of motion for the quantum field
$\phi$
\begin{proposition}
The quantum field $\phi(x)$ satisfies
\be
\square \phi(x) = -\frac 1{4\pi{\rm R}^2} \phi(v_0)\ , \label{eq:eqofmot}
\ee
where $v_0$ is the previously defined state represented by $v_0 =
[-4\pi\mathrm R^2\square h]$.
\end{proposition}
{\bf Proof}.
For any test function $f$ we have $f= f_0 + h\int f$ the decomposition
in its $\mathcal D_0$ and $H$ components. We
can obviously write $\phi(f)=\phi(f_0)+\int f\phi(h)$.
Recalling that $\square \phi_{|_{\mathcal D_0}}\equiv 0$, we have
\beas
&& <\square\phi(f)\phi(g)> = \int f <\phi(\square h)\phi(g)> = \\
&&= -\frac
1{4\pi {\rm R}^2}\int f \int g =: -\frac
1{4\pi {\rm R}^2}  \int f <\phi(v_0)\phi(g)> \ .
\eeas
This, together with the fact that $\phi$ is a free field, proves the
statement. Q.E.D.\par\vskip 3pt
In the homogeneous case the equation $\square \phi = 0$ is  clearly
invariant under the transformation
$\phi\to\phi+\lambda{\mathbf 1}$. This invariance is still present as
the following Lemma shows
\begin{lemma}
The equation of motion for the quantum field $\phi$ is invariant under
 the {\em gauge transformation} $\gamma^\lambda(\phi):= \phi +
 \lambda$, $\lambda   \in \mathbb C$.
\end{lemma}
{\bf Proof}.
The
 identity operator
$\mathbf 1$, as an operator-valued distribution, associates to a test
function its total integral.  In the equation of motion we then have
\bes
\square \gamma^\lambda(\phi) =\square \phi = -\frac 1{4\pi \mathrm
R^2} \phi(v_0) =  -\frac 1{4\pi \mathrm
R^2} \gamma^\lambda(\phi) (v_0)\ ,
\ees
where the last equality follows from the fact that the state $v_0=[4\pi
\mathrm R^2 \square h]$ has zero total integral. Q.E.D.\par\vskip 3pt
 We can then define
a {\em gauge transformation} as the automorphism of the field algebra
generated by
\be
\gamma^\lambda:\phi\to\phi+\lambda,\quad \lambda\in{\mathbb C}.
\ee
Let us now introduce the following operator 
\be
 Q=
i\left[\phi^+(v_0)-\phi^-(v_0)\ri], 
\ee
 where  the operators
$\phi^\pm(f)$ are defined as creators--annihilators of states $v_0$
in the Fock-Krein space $\mathcal K$. This is a continuous operator
from the $n$-particle space ${\mathcal K}^{(n)}$ to ${\mathcal
K}^{(n\pm1)}$. Moreover it is easily verified that $Q$ satisfies the
following commutation relations: 
\be
 \left[Q,\phi(f)\ri]=-i\int{\rm
d}\sigma(x)f(x). \label{equ:commq} 
\ee
 One can compute directly that
--exactly as in the flat case-- \bea
 [\phi_\pm(v_0), \phi(f)] = \mp \frac 1 2 \int{\rm d}\sigma(x)
f(x)\ ;\qquad
 [\phi_+(v_0),\phi_-(v_0)] = 0\ .\label{eqs:infracomm}
\eea
The natural question now arises as to whether  this {\em
charge} $Q$ is the integral of a local expression.
The classical charge of any solution of the wave equation in $d+1$
dimensions $
\square \varphi=0 $ is defined by
integrating along any spacelike $d$-surface the ``timelike''
derivative along the direction ortogonal to the surface
\be
\int_\Sigma {\rm d}v_\Sigma \pa_{\hat n} \varphi.\label{69a}
\ee
Such expression is ``conserved'' (i.e. independent of the spacelike
surface $\Sigma$) because of the equation of motion.
The integrand in eq. (\ref{69a}) is the Hodge dual of
the $1$-form ${\rm d}\varphi$, i.e. (in Lorentzian signature) the
$d$--form defined by
\be
({\rm d}\varphi)^* :=
\sqrt{-g}\epsilon_{\mu_1,...,\mu_d,\nu}g^{\nu\rho}\pa_\rho \varphi
{\rm d}x^{\mu_1}\cdots {\rm d}x^{\mu_d}\ .
\ee
Such a $d$-form can be integrated on any $d$-surface in an intrinsic
way.
Now, such a form is {\em closed} iff the function $\varphi$ satisfies
the wave equation $\square \varphi=0$ because
\bes
{\rm d} ({\rm d}\varphi)^* \propto  \square \varphi\, {\rm d}vol\ .
\ees
The integral of a closed $d$-form on a
$d$-surface is independent of continuous deformations of the surface
(i.e. the choice of the ``time-slice'' in our setting).\par
This preamble shows clearly that we are to expect some problems from
the fact that our quantum field $\phi$ does not satisfy the
homogeneous wave equation.\\
Let us have a closer look at what happens in the case at hand:
the explicit expression of the Hodge dual is now the $d=1$-form  (see
also later the discussion about the dual field)
\bes
({\rm d}\phi)^* = -\pa_\tau \phi {\rm d}\theta
-\pa_\theta\phi{\rm d} \tau
\ees
for our (quantum) field $\phi$ (in conformal coordinates for clarity).\\
The equation of motion  (\ref{eq:eqofmot}) implies that this
operator--valued form is
not closed
\bes
{\rm d} ({\rm d}\phi)^*  = \frac 1{4\pi\mathrm R^2} \phi(v_0)\, {\rm
d}vol = \frac 1{4\pi\mathrm R^2} \phi(v_0) \frac{ {\rm
d}\tau\wedge {\rm d}\theta}{\cos^2(\tau/\mathrm R)} \ .
\ees
Therefore the charge defined as in the flat case
 (with all the
additional technical details which we  now omit) \cite{[MPS],[MPS2]} by
\bes
Q \propto  \int {\rm d}x^1\, \pa_{x^0} \phi
\ees
could not be possibly conserved (in our case the expression would look more like
 $\int {\rm d}\theta\, \pa_\tau \phi$).
\begin{proposition}
The charge $Q= \frac i 2\left(\phi^{+}(v_0)-\phi^{-}(v_0)\ri)$ is
the integral of a local current, namely 
\be
 Q = -\frac 1{8\pi}
\int_{\Sigma} \left( ({\rm d} \phi)^* + \frac 1{4\pi {\rm R}}
\tan(\tau/{\rm R}) \phi(v_0) {\rm d}\theta \ri)\ ,
\label{eq:localcharge}
 \ee
  where $\Sigma$ is any spacelike
$d=1$--surface (i.e. a curve, e.g. $\tau=const$). The integral is
independent of the spacelike (closed, simple) curve $\Sigma$ because
the one--form in the integrand is closed.
\end{proposition}
\begin{remark}
The integrand of Eq. \ref{eq:localcharge} is the differential
of a  {\bf dual field} $\widetilde{\phi}$ up to an exact form: indeed we
 {\em define} the dual field by the equation
\be
 {\rm d}\widetilde{\phi} = ({\rm d}\phi)^* - \frac {
\tan(\tau/{\rm R})}{4\pi{\rm R}} \left(\phi(v_0) {\rm d}\theta
+\widetilde  \phi(v_0) {\rm d}\tau\ri)\ , 
\ee
 which differs from the
integrand in Eq. \ref{eq:localcharge} by the exact operator--valued
form 
\be
 \frac { \tan(\tau/{\rm R})}{4\pi{\rm R}}
\widetilde{\phi}(v_0){\rm d}\tau =\frac 1{4\pi} \widetilde\phi(v_0)
{\rm d} \ln(\cos(\tau/{\rm R}))\ ,
 \ee
  which does not contribute to
the integral. In this notation the above proposition would read 
\be
Q =-\frac 1{8\pi}  \int_\Sigma {\rm d}\widetilde \phi\ . 
\ee
 The
one--form ${\rm d}\widetilde\phi$ is (classically) closed but not
exact, i.e. the dual field $\widetilde \phi$ naturally lives on the
universal covering of $X_2$. In other words\
 the dual field $\widetilde \phi$
carries a topological charge w.r.t. $\phi$ (see later).
\end{remark}
{\bf Sketch of Proof}.\\
The proof consists in showing that the integrand is indeed a closed
form (which is straightforward) and that the commutator of the
integral with the fields $\phi(f)$ reproduces the correct commutator
of $Q$ as given in (\ref{equ:commq}), which amounts to a direct
manipulation of the integrals. Q.E.D.
From the relation 
\be 
\frac{{\rm d}}{{\rm
d}\lambda}\gamma^\lambda(\phi(f))=\int_{X_2}{\rm
d}\sigma(x)f(x)=i\left[Q,\phi(f)\ri]
 \ee
  follows that the
automorphism $\gamma^\lambda$ is generated by the operator $Q$.
\begin{theorem}
The automorphism $\gamma^\lambda$ is implementable in the Krein space ${\mathcal K}$ by the
$\eta$-unitary operator
\be
\gamma^\lambda=e^{i\lambda Q}.
\ee
\end{theorem}
The not difficult proof will be detailed elsewhere.

The Wightman function ${\mathcal W}_0(x,x')$ is not invariant under
the gauge transformation:
indeed, for $\phi\to\phi+\lambda$, we get
 ${\mathcal W}_0\to{\mathcal W}_0+\lambda^2$. For this
reason some authors (\cite{[DBR],[GRT]}) do not use Witghman formalism
to construct a QFT  for
the massless minimally
coupled field in de Sitter space--time. The construction is nonetheless still
feasible because the
vacuum is not gauge invariant but it is mapped to a non-physical
state ({\em i.e.} a zero--norm
state). If restricted to the physical space of states
 ${\mathcal H}_{phys}$ the two point
function is positive, analytic and invariant under de Sitter and gauge
transformation.
%
%%%%%%%%%%%%%%%%%%%%%%%%%%%conclusioni%%%%%%%%%%%%%%%%%%%%%%%%%%%%%%%
\section{Conclusions}
We have constructed a QFT of a massless minimally coupled scalar field in a bidimensional de
Sitter space--time. Although we have studied this particular case, the results are valid also
for
the four dimensional de Sitter universe. The renormalized two point function for a massless
minimally coupled scalar fields in $X_4$ isn't positive defined and one can construct a full
de Sitter invariant vacuum with the same techniques we have described in section
\ref{sec:krein}.

This seems to disagree with Allen's theorem (\cite{[A]}) which states that in a four
dimensional
de Sitter space time can't exist a de Sitter invariant vacuum for a massless scalar field.
As we have already pointed out, the
problem is analogous to the case of the vacuum of the massless scalar field in a bidimensional
Minkowski space--time: it is impossible to find a two point function positive defined,
analytic, Lorentz invariant and local (\cite{[S]}); if one insists over Lorentz
invariance necessarily loose positivity.

If one use the normal modes to construct the two point function, its positivity is a natural
consequence of the construction itself and, in agreement with Allen's theorem, de Sitter
invariance is lost. With the normal modes construction, not only dS invariance but also
gauge invariance is lost, being the Hilbert space of the states non--invariant under gauge
transformation.

Another result is that, being the vacuum de Sitter invariant, the linear dependence by the
time coordinate of the vacuum fluctuation for the non--invariant vacuum is lost
(\cite{[A],[AF]}).
 This result could have some consequence for
the inflationary model, where this time dependence is used to explain the roll--over of the
inflation fields responsible of the inflation. It is important to point out that this results
are valid only for a de Sitter universe that exist forever and we don't yet know which are the
implication for a de Sitter universe with a finite life.
%%%%%%%%%%%%%%%%%%%%%%%%%%%%%%%%%%%%%%%%%%%%%%%%%%%%%%%%%%%%%%%%%%%%%%%%%%%%%%%%%%%%%%%%%%

\end{document}